\documentclass[aps,prb,superscriptaddress,preprint,showpacs,preprintnumbers,amsmath,amssymb]{revtex4}

\usepackage{graphicx}
\usepackage{dcolumn}
\usepackage{multirow}
\usepackage{bm}
\usepackage{mathrsfs}

\newcommand{\mytensor}[1]{\mathord{\buildrel{\lower3pt\hbox{$\scriptscriptstyle\leftrightarrow$}}\over{#1}}}

\begin{document}
\title{Non-local effective medium of metamaterial}
\author{Jensen Li}
\author{J. B. Pendry}
\affiliation{Blackett Laboratory, Imperial College, Prince Consort Road, London SW7 2BW, United Kingdom}

\begin{abstract}
In this paper, we investigate the effect of spatial dispersion on a 
double-lattice metamaterial which has both magnetic and electric response. A 
numerical scheme based on a dipolar model is developed to extract the 
non-local effective medium of the metamaterial. We found a 
structure-induced bianisotropy near the resonance gap even if the artificial 
particles are free from cross-coupling. A cross-coupled resonance also 
results from spatial dispersion and it can be understood 
using a Lorentz frequency dispersion model. The strength of the cross-coupled 
resonance depends on the microstructure and it is larger at a higher normalized frequency.
\end{abstract}\date{\today}


\maketitle

\section{Introduction}

Metamaterials consisting arrays of artificial particles 
usually work in the frequency regime in which the free-space wavelength 
is larger than the inter-particle distance by a factor around 10.
Their electromagnetic properties can therefore be understood by treating 
a metamaterial as a homogeneous medium.
More often, a local effective medium of permittivity 
($\epsilon_{eff}$) and permeability ($\mu_{eff}$) is assumed to be valid.
There are several approaches in obtaining the local effective medium of 
a metamaterial. One approach is the S-parameter retrieval method. \cite{Smith:2005} 
It obtains the material parameters by comparing the transmission and the reflection coefficients of a slab of metamaterial 
and the corresponding homogeneous slab. Another approach relies on a homogenization theory 
which averages the microscopic field by line and area integrals. \cite{Pendry:1999,Smith:2006,Lerat:2006}
Both approaches can cater for complex geometry and microstructure of the metamaterial and they 
work very well at frequencies far away from the resonance gap where spatial dispersion becomes 
important so that higher order eigenmodes (including the non-propagating ones) can be excited inside the 
metamaterial. \cite{Belov:2006} In this work, we will take another approach to deal with the spatial dispersion at a frequency near the 
resonance gap.
Instead of picking or assuming a dominant eigenmode within the metamaterial, we will find 
the non-local effective medium directly for an arbitrary Bloch wave vector $\bf q$. For instance, 
the permittivity $\epsilon(\omega,{\bf q})$ now depends on both the angular frequency $\omega$ and the wave vector.

Apart from working near the resonance gap, some metamaterials like the wire medium 
, \cite{Belov:2003,Silveirinha:2006} the three-dimensional wire mesh \cite{Shapiro:2006} and the structured metal 
surface \cite{Garc:2005} also indicate that spatial dispersion (the non-local 
effective medium) can become important in order to have an accurate description of the metamaterial. In this paper, 
we will establish a numerical scheme in getting the non-local effective medium and some effects of spatial 
dispersion on metamaterial will be examined.

Section \ref{sect2} outlines the numerical scheme in obtaining the non-local 
effective medium. We will employ a dipolar model for simplicity. It has the advantage that 
analytic formulas can be obtained in some cases. Moreover, all the material parameters, two 
tensors ($\mytensor \epsilon$ and $\mytensor \mu$) together with two pseudo tensors ($\mytensor \xi$ and $\mytensor \zeta$) can be obtained 
easily in the whole frequency regime. All the four constitutive tensors will be extracted as 
Ref. \onlinecite{Marques:2002} and \onlinecite{Xudong:2005} pointed 
out that a metamaterial should be described as a bianisotropic medium if the artificial atoms suffer from 
cross-coupling (electric field generates magnetic dipole and/or magnetic field generates electric dipole).

In Section \ref{sect3}, the local effective medium of a double-lattice metamaterial will be first examined. 
One sublattice consists of electric artificial atoms and another sublattice consists of magnetic artificial 
atoms. The condition that the local effective medium becoming free of bianisotropy will be discussed.

In the last section, the non-local effective medium of the double-lattice metamaterial will be examined. 
The relationship between the non-local effective medium to the corresponding local one will be discussed. 
On the other hand, a Lorentz-type model is suggested for fitting the frequency dispersion of the non-local 
effective medium.
Some effects of the spatial dispersion, including the structure-induced bianisotropy and the cross-coupled 
resonance, will also be discussed.

\section{Formulation}
\label{sect2}

\noindent
Here, we give an outline of the numerical scheme in obtaining the non-local 
effective medium using a dipolar model. The artificial particles ( 
split-rings, short wires, or electric atoms, \cite{Schurig:2006} etc.) 
are placed in an infinite lattice of lattice vectors ${\rm {\bf R}}$ in 
vacuum (of wavenumber $k_0 $, wave speed $c_0 $, and intrinsic impedance 
$\eta _0 )$. In each unit cell labeled by ${\rm {\bf R}}$, there are $n$ 
particles (indexed by $j = 1,...,n)$ at positions ${\rm {\bf R}} + {\bf \tau}_j $. 
In the dipolar limit, each artificial particle has an electric 
dipole moment ${\rm {\bf p}}\left( {{\rm {\bf R}} + {\rm {\bf \tau }}_j } 
\right)$ and a magnetic dipole moment ${\rm {\bf m}}\left( {{\rm {\bf R}} + 
{\rm {\bf \tau }}_j } \right)$ and they are governed by the following 
equation (with time dependence factor $e^{ - i\omega t}$ and S. I. units):

\begin{equation}
\label{eq1}
\left[ {{\begin{array}{*{20}c}
 {c_0 {\rm {\bf p}}\left( {{\rm {\bf R}} + {\rm {\bf \tau }}_j } \right)} 
\hfill \\
 {{\rm {\bf m}}\left( {{\rm {\bf R}} + {\rm {\bf \tau }}_j } \right)} \hfill 
\\
\end{array} }} \right] = {\alpha }_j \cdot \left[ {{\begin{array}{*{20}c}
 {{\rm {\bf E}}_{\mbox{local}} \left( {{\rm {\bf R}} + 
{\rm {\bf \tau }}_j } \right) / \eta _0 } \hfill \\
 {{\rm {\bf H}}_{\mbox{local}} \left( {{\rm {\bf R}} + 
{\rm {\bf \tau }}_j } \right)} \hfill \\
\end{array} }} \right],
\end{equation}

\noindent
where ${{\rm {\bf E}}_{\mbox{local}} \left( {{\rm {\bf R}} + {\rm {\bf \tau }}_j } \right)} $ 
and ${{\rm {\bf H}}_{\mbox{local}} \left( {{\rm {\bf R}} + {\rm {\bf \tau }}_j } \right)} $ 
are the local electric and local 
magnetic fields at the center of the $j$-th particle. These local fields 
include the fields radiated by the external sources and the scattered fields 
from all the other particles. The 6-by-6 matrix ${\alpha }_j $ represents 
the polarizability of the $j$-th particle.

In this model, every artificial particle is replaced by an equivalent 
dipolar point particle. In order to extract the non-local effective medium of 
this crystal at an arbitrary wave vector ${\rm {\bf q}}$, we drive the 
crystal by the external fields in the form

\begin{equation}
\label{eq2}
\begin{split}
 {\rm {\bf E}}_{\mbox{ext}} \left( {\rm {\bf r}} \right) & = {\rm {\bf 
E}}_{\mbox{ext}} \exp \left( {i{\rm {\bf q}} \cdot {\rm {\bf r}}} \right), 
\\ 
 {\rm {\bf H}}_{\mbox{ext}} \left( {\rm {\bf r}} \right) & = {\rm {\bf 
H}}_{\mbox{ext}} \exp \left( {i{\rm {\bf q}} \cdot {\rm {\bf r}}} \right). 
\\ 
 \end{split}
\end{equation}

\noindent
These external fields are in fact generated by an electric and a magnetic 
polarization of the same spatial dependence. The crystal responds to it by 
generating dipole moments in the form

\begin{equation}
\label{eq3}
\left[ {{\begin{array}{*{20}c}
 {c_0 {\rm {\bf p}}\left( {{\rm {\bf R}} + {\rm {\bf \tau }}_j } \right)} 
\hfill \\
 {{\rm {\bf m}}\left( {{\rm {\bf R}} + {\rm {\bf \tau }}_j } \right)} \hfill 
\\
\end{array} }} \right] = \left[ {{\begin{array}{*{20}c}
 {c_0 {\rm {\bf p}}_j } \hfill \\
 {{\rm {\bf m}}_j } \hfill \\
\end{array} }} \right]e^{i{\rm {\bf q}} \cdot {\rm {\bf R}}},
\end{equation}

\noindent
which satisfies the multiple-scattering equation:

\begin{equation}
\label{eq4}
\sum\limits_{j' = 1}^n {\left( {{T}^{ - 1}} \right)_{jj'} \left( {\rm {\bf q}} 
\right)} \cdot \frac{1}{V}\left[ {{\begin{array}{*{20}c}
 {c_0 {\rm {\bf p}}_{j'} } \hfill \\
 {{\rm {\bf m}}_{j'} } \hfill \\
\end{array} }} \right]e^{ - i{\rm {\bf q}} \cdot {\rm {\bf \tau }}_{j'} } = 
\left[ {{\begin{array}{*{20}c}
 {{\rm {\bf E}}_{\mbox{ext}} / \eta _0 } \hfill \\
 {{\rm {\bf H}}_{\mbox{ext}} } \hfill \\
\end{array} }} \right],
\end{equation}

\noindent
with $V$ being the volume of a single primitive unit cell.
The $6n\times 6n$ transition matrix ${T}$ is given by its inverse:

\begin{equation}
\label{eq5}
\left( {{T}^{ - 1}} \right)_{jj'} \left( {\rm {\bf q}} \right) = V{\alpha 
}_j^{ - 1} \delta _{jj'} - {G}_{jj'} \left( {\rm {\bf q}} \right),
\end{equation}

\noindent
where the lattice Green's function ${G}_{jj'} $ is defined by

\begin{equation}
\label{eq6}
{G}_{jj'} \left( {\rm {\bf q}} \right) = V\sum\limits_{{\rm {\bf r}} = {\rm 
{\bf R}} + {\rm {\bf \tau }}_j - {\rm {\bf \tau }}_{j'} \ne {\rm {\bf 0}}} {e^{ 
- i{\rm {\bf q}} \cdot {\rm {\bf r}}}{G}_0 \left( {\rm {\bf r}} \right)} .
\end{equation}

\noindent
The 6-by-6 matrix ${G}_0 \left( {\rm {\bf r}} \right)$ which relates a dipole 
moment to its radiation field, together with its Fourier transform ${G}_0 
\left( {\rm {\bf q}} \right)$ are given explicitly in Appendix \ref{appendixa}. The 
lattice Green's function can be evaluated by the Ewald sum technique. For 
mathematical convenience, we also define

\begin{equation}
\label{eq7}
{G}_{jj'} \left( {\rm {\bf q}} \right) + \frac{ik_0^3 V}{6\pi }{I}\delta 
_{jj'} = {G}_0 \left( {\rm {\bf q}} \right) + \left[ {{\begin{array}{*{20}c}
 
{\mathord{\buildrel{\lower3pt\hbox{$\scriptscriptstyle\leftrightarrow$}}\over 
{\beta }} _{jj'} \left( {\rm {\bf q}} \right)} \hfill & 
{\mathord{\buildrel{\lower3pt\hbox{$\scriptscriptstyle\leftrightarrow$}}\over 
{\gamma }} _{jj'} \left( {\rm {\bf q}} \right)} \hfill \\
 { - 
\mathord{\buildrel{\lower3pt\hbox{$\scriptscriptstyle\leftrightarrow$}}\over 
{\gamma }} _{jj'} \left( {\rm {\bf q}} \right)} \hfill & 
{\mathord{\buildrel{\lower3pt\hbox{$\scriptscriptstyle\leftrightarrow$}}\over 
{\beta }} _{jj'} \left( {\rm {\bf q}} \right)} \hfill \\
\end{array} }} \right],
\end{equation}

\noindent
where ${I}$ is the $6\times 6$ Identity Matrix. The tensors 
$\mathord{\buildrel{\lower3pt\hbox{$\scriptscriptstyle\leftrightarrow$}}\over 
{\beta }} _{jj'} \left( {\rm {\bf q}} \right)$ and 
$\mathord{\buildrel{\lower3pt\hbox{$\scriptscriptstyle\leftrightarrow$}}\over 
{\gamma }} _{jj'} \left( {\rm {\bf q}} \right)$ are periodic functions in the 
reciprocal space. They can be characterized according to the point group \cite{Dmitriev:2000}
associated with the lattice structure at a particular wave vector ${\rm {\bf 
q}}$ together with the requirement that the matrix transpose (fixed $j$ and $j'$) of 
$\mathord{\buildrel{\lower3pt\hbox{$\scriptscriptstyle\leftrightarrow$}}\over 
{\gamma }} _{jj'} \left( {\rm {\bf q}} \right)$ satisfies

\begin{equation}
\label{eq8}
\mathord{\buildrel{\lower3pt\hbox{$\scriptscriptstyle\leftrightarrow$}}\over 
{\gamma }} _{jj'} \left( {\rm {\bf q}} \right)^T = - 
\mathord{\buildrel{\lower3pt\hbox{$\scriptscriptstyle\leftrightarrow$}}\over 
{\gamma }} _{jj'} \left( {\rm {\bf q}} \right),
\end{equation}

\noindent
which is already implied from Eq. (\ref{eq6}) with Eq. (\ref{eqa2}) . 
Table \ref{tab1} lists the forms of the tensors 
$\mathord{\buildrel{\lower3pt\hbox{$\scriptscriptstyle\leftrightarrow$}}\over 
{\beta }} _{jj'} \left( {\rm {\bf q}} \right)$ and 
$\mathord{\buildrel{\lower3pt\hbox{$\scriptscriptstyle\leftrightarrow$}}\over 
{\gamma }} _{jj'} \left( {\rm {\bf q}} \right)$ for some common single 
lattices and double lattices. The matrices are listed with respect to the 
Cartesian basis vectors $\left\{ {\hat {x},\hat {y},\hat {z}} \right\}$ 
where $\hat {z}$ is defined to be the direction along ${\rm {\bf q}}$ in our 
convention.

\begin{table}[htbp]
\begin{tabular}
{|m{55pt}|m{50pt}|c|c|}
\hline
& 
${\rm {\bf q}}$& 
$\mathord{\buildrel{\lower3pt\hbox{$\scriptscriptstyle\leftrightarrow$}}\over {\beta }} _{jj'} \left( {\rm {\bf q}} \right)$& 
$\mathord{\buildrel{\lower3pt\hbox{$\scriptscriptstyle\leftrightarrow$}}\over {\gamma }} _{jj'} \left( {\rm {\bf q}} \right)$ \\
\hline
\multirow{4}{55pt}{SC/ FCC/ BCC single lattice structure} &
At $\Gamma $& 
$\beta $& 
0 \\
\cline{2-4} 
 & 
Along [111]/[001] &
$\left[ {{\begin{array}{ccc}
 {\beta _T } \hfill & 0 \hfill & 0 \hfill \\
 0 \hfill & {\beta _T } \hfill & 0 \hfill \\
 0 \hfill & 0 \hfill & {\beta _L } \hfill \\
\end{array} }} \right]$& 
$\left[ {{\begin{array}{*{20}c}
 0 \hfill & { - \gamma } \hfill & 0 \hfill \\
 \gamma \hfill & 0 \hfill & 0 \hfill \\
 0 \hfill & 0 \hfill & 0 \hfill \\
\end{array} }} \right]$ \\
\hline
NaCl/ CsCl &
At $\Gamma $& 
$\beta $& 
0 \\
\cline{2-4} 
 & 
Along [111]/[001]& 
$\left[ {{\begin{array}{*{20}c}
 {\beta _T } \hfill & 0 \hfill & 0 \hfill \\
 0 \hfill & {\beta _T } \hfill & 0 \hfill \\
 0 \hfill & 0 \hfill & {\beta _L } \hfill \\
\end{array} }} \right]$& 
$\left[ {{\begin{array}{*{20}c}
 0 \hfill & { - \gamma } \hfill & 0 \hfill \\
 \gamma \hfill & 0 \hfill & 0 \hfill \\
 0 \hfill & 0 \hfill & 0 \hfill \\
\end{array} }} \right]$ \\
\hline
\multirow{4}{60pt}{Zn Blende} &
At $\Gamma $& 
$\beta $& 
0 \\
\cline{2-4} 
 & 
Along [111]& 
$\left[ {{\begin{array}{*{20}c}
 {\beta _T } \hfill & 0 \hfill & 0 \hfill \\
 0 \hfill & {\beta _T } \hfill & 0 \hfill \\
 0 \hfill & 0 \hfill & {\beta _L } \hfill \\
\end{array} }} \right]$& 
$\left[ {{\begin{array}{*{20}c}
 0 \hfill & { - \gamma } \hfill & 0 \hfill \\
 \gamma \hfill & 0 \hfill & 0 \hfill \\
 0 \hfill & 0 \hfill & 0 \hfill \\
\end{array} }} \right]$ \\
\cline{2-4} 
 &
Along [001]: $\hat {x}\parallel \left[ {1 1 0} \right]$, $\hat {y}\parallel \left[ { -1 1 0} \right]$& 
$\left[ {{\begin{array}{*{20}c}
 {\beta _{xx} } \hfill & 0 \hfill & 0 \hfill \\
 0 \hfill & {\beta _{yy} } \hfill & 0 \hfill \\
 0 \hfill & 0 \hfill & {\beta _L } \hfill \\
\end{array} }} \right]$& 
$\left[ {{\begin{array}{*{20}c}
 0 \hfill & { - \gamma } \hfill & 0 \hfill \\
 \gamma \hfill & 0 \hfill & 0 \hfill \\
 0 \hfill & 0 \hfill & 0 \hfill \\
\end{array} }} \right]$ \\
\hline
\end{tabular}
\caption{The tensor forms of the lattice Green's function for different single and 
double lattice structures along different directions of the wave vector.}
\label{tab1}
\end{table}

In the next step, the dipole moment distributions numerically solved from Eq. (\ref{eq4}) 
are then macroscopically averaged. The procedure we choose here is to apply a low pass filter 
such that all the Fourier components outside the first Brillouin zone are eliminated. 
In essence, it means that all the microscopic fields are ensemble averaged by translating the whole crystal with 
an arbitrary distance while the external fields remain unchanged.
Thus, the macroscopic polarizations are written as

\begin{equation}
\label{eq9}
\begin{split}
 c_0 \left\langle {\rm {\bf P}} \right\rangle \left( {\rm {\bf r}} \right) & = 
c_0 \left\langle {\rm {\bf P}} \right\rangle e^{i{\rm {\bf q}} \cdot {\rm 
{\bf r}}} = \frac{1}{V}\sum\limits_{j = 1}^n {c_0 {\rm {\bf p}}_j e^{ - 
i{\rm {\bf q}} \cdot {\rm {\bf \tau }}_j }} e^{i{\rm {\bf q}} \cdot {\rm 
{\bf r}}}, \\ 
 \left\langle {\rm {\bf M}} \right\rangle \left( {\rm {\bf r}} \right) & = 
\left\langle {\rm {\bf M}} \right\rangle e^{i{\rm {\bf q}} \cdot {\rm {\bf 
r}}} = \frac{1}{V}\sum\limits_{j = 1}^n {{\rm {\bf m}}_j e^{ - i{\rm {\bf 
q}} \cdot {\rm {\bf \tau }}_j }} e^{i{\rm {\bf q}} \cdot {\rm {\bf r}}}. \\ 
 \end{split}
\end{equation}

\noindent
The macroscopic polarizations can now be formally written as

\begin{equation}
\label{eq10}
\left[ {{\begin{array}{*{20}c}
 {c_0 \left\langle {\rm {\bf P}} \right\rangle } \hfill \\
 {\left\langle {\rm {\bf M}} \right\rangle } \hfill \\
\end{array} }} \right] = \left\langle {T} \right\rangle \left( {\rm {\bf q}} 
\right)\left[ {{\begin{array}{*{20}c}
 {{\rm {\bf E}}_{\mbox{ext}} / \eta _0 } \hfill \\
 {{\rm {\bf H}}_{\mbox{ext}} } \hfill \\
\end{array} }} \right],
\end{equation}

\noindent
where the averaged transition matrix is given by

\begin{equation}
\label{eq11}
\left\langle {T} \right\rangle \left( {\rm {\bf q}} \right) = 
\sum\limits_{j,{j'} = 1}^n {{T}_{jj'} \left( {\rm {\bf q}} \right)} .
\end{equation}

\noindent
Eq. (\ref{eq10}) summarizes the response of the crystal to the external fields. 
Then, the macroscopic fields are constructed from the sum of the external 
fields and the secondary radiation by

\begin{equation}
\label{eq12}
\left[ {{\begin{array}{*{20}c}
 {\left\langle {\rm {\bf E}} \right\rangle / \eta _0 } \hfill \\
 {\left\langle {\rm {\bf H}} \right\rangle } \hfill \\
\end{array} }} \right] = {G}_0 \left( {\rm {\bf q}} \right) \cdot \left[ 
{{\begin{array}{*{20}c}
 {c_0 \left\langle {\rm {\bf P}} \right\rangle } \hfill \\
 {\left\langle {\rm {\bf M}} \right\rangle } \hfill \\
\end{array} }} \right] + \left[ {{\begin{array}{*{20}c}
 {{\rm {\bf E}}_{\mbox{ext}} / \eta _0 } \hfill \\
 {{\rm {\bf H}}_{\mbox{ext}} } \hfill \\
\end{array} }} \right].
\end{equation}

\noindent
Finally, by combining Eq. (\ref{eq10}) and Eq. (\ref{eq12}), we obtain the constitutive 
relationship

\begin{equation}
\label{eq13}
\left[ {{\begin{array}{*{20}c}
 {c_0 \left\langle {\rm {\bf P}} \right\rangle } \hfill \\
 {\left\langle {\rm {\bf M}} \right\rangle } \hfill \\
\end{array} }} \right] = 
{
\left[ {{\begin{array}{*{20}c}
{\mathord{\buildrel{\lower3pt\hbox{$\scriptscriptstyle\leftrightarrow$}}\over 
{\epsilon }} \left( {\rm {\bf q}} \right) - 
\mathord{\buildrel{\lower3pt\hbox{$\scriptscriptstyle\leftrightarrow$}}\over 
{I}} } \hfill & 
{\mathord{\buildrel{\lower3pt\hbox{$\scriptscriptstyle\leftrightarrow$}}\over 
{\xi }} \left( {\rm {\bf q}} \right)} \hfill \\
{\mathord{\buildrel{\lower3pt\hbox{$\scriptscriptstyle\leftrightarrow$}}\over 
{\zeta }} \left( {\rm {\bf q}} \right)} \hfill & 
{\mathord{\buildrel{\lower3pt\hbox{$\scriptscriptstyle\leftrightarrow$}}\over 
{\mu }} \left( {\rm {\bf q}} \right) - 
\mathord{\buildrel{\lower3pt\hbox{$\scriptscriptstyle\leftrightarrow$}}\over 
{I}} } \hfill \\
\end{array} }} \right]
}
 \cdot \left[ 
{{\begin{array}{*{20}c}
 {\left\langle {\rm {\bf E}} \right\rangle / \eta _0 } \hfill \\
 {\left\langle {\rm {\bf H}} \right\rangle } \hfill \\
\end{array} }} \right],
\end{equation}

\noindent
where the four constitutive tensors are given by

\begin{equation}
\label{eq14}
\left[ {{\begin{array}{*{20}c}
{\mathord{\buildrel{\lower3pt\hbox{$\scriptscriptstyle\leftrightarrow$}}\over 
{\epsilon }} \left( {\rm {\bf q}} \right) - 
\mathord{\buildrel{\lower3pt\hbox{$\scriptscriptstyle\leftrightarrow$}}\over 
{I}} } \hfill & 
{\mathord{\buildrel{\lower3pt\hbox{$\scriptscriptstyle\leftrightarrow$}}\over 
{\xi }} \left( {\rm {\bf q}} \right)} \hfill \\
{\mathord{\buildrel{\lower3pt\hbox{$\scriptscriptstyle\leftrightarrow$}}\over 
{\zeta }} \left( {\rm {\bf q}} \right)} \hfill & 
{\mathord{\buildrel{\lower3pt\hbox{$\scriptscriptstyle\leftrightarrow$}}\over 
{\mu }} \left( {\rm {\bf q}} \right) - 
\mathord{\buildrel{\lower3pt\hbox{$\scriptscriptstyle\leftrightarrow$}}\over 
{I}} } \hfill \\
\end{array} }} \right]
^{ - 1} = \left\langle {T} \right\rangle ^{ - 1}\left( 
{\rm {\bf q}} \right) + {G}_0 \left( {\rm {\bf q}} \right).
\end{equation}

\noindent
In general, 
apart from the permittivity tensor 
$\mathord{\buildrel{\lower3pt\hbox{$\scriptscriptstyle\leftrightarrow$}}\over 
{\epsilon }} \left( {\rm {\bf q}} \right)$ and the permeability tensor 
$\mathord{\buildrel{\lower3pt\hbox{$\scriptscriptstyle\leftrightarrow$}}\over 
{\mu }} \left( {\rm {\bf q}} \right)$, we also obtain the off-diagonal 
tensors 
$\mathord{\buildrel{\lower3pt\hbox{$\scriptscriptstyle\leftrightarrow$}}\over 
{\xi }} \left( {\rm {\bf q}} \right)$ and 
$\mathord{\buildrel{\lower3pt\hbox{$\scriptscriptstyle\leftrightarrow$}}\over 
{\zeta }} \left( {\rm {\bf q}} \right)$ which represent the bianisotropy of 
the metamaterial. We will see in the next section that the bianisotopy 
should not be neglected in considering the non-local effective medium. All 
the four tensors are functions of the frequency $\omega $ and the wave 
vector ${\rm {\bf q}}$. For convenience, we will omit writing 
$\omega $ inside the parenthesis for the constitutive tensors. Moreover, we 
will further omit the dependence of ${\rm {\bf q}}$ if the local limit ${\rm 
{\bf q}} = {\rm {\bf 0}}$ is taken, i.e. 
$\mathord{\buildrel{\lower3pt\hbox{$\scriptscriptstyle\leftrightarrow$}}\over 
{\epsilon }} \left( {{\rm {\bf q}} = {\rm {\bf 0}}} \right)$ is 
abbreviated as 
$\mathord{\buildrel{\lower3pt\hbox{$\scriptscriptstyle\leftrightarrow$}}\over 
{\epsilon }} $. 

We note that in our scheme of defining the non-local effective medium 
at a particular ${\rm {\bf q}}$, a fixed form of external fields (Eq. (\ref{eq2})) is used. 
There are alternative methods which directly deal with the macroscopic 
fields without considering the external fields. \cite{Ponti:2001,Ponti:2002}

\section{The local effective medium - double lattice}
\label{sect3}

\noindent
In this work, we will concentrate on the double lattice structure which is a 
common type of microstructures for metamaterials in obtaining a negative 
refractive index. The double lattice consists of two types of artificial 
particles. One type of the particles (labeled by $j = 1)$ has a 
resonating electric response (e.g. a short wire) and another (labeled by 
$j = 2)$ has a resonating magnetic response (e.g. a split-ring). In 
particular, we would like to study the importance of the electro-magnetic 
coupling between the electric and the magnetic particles when we have a 
crystal structure having both electric and magnetic response. It is a 
general feature of metamaterials.

First, we would like to restrict our discussion to the class of 
metamaterials whose local effective medium is free from bianisotropy. It can 
be achieved through a careful design of the artificial particles such that 
the particles are free from cross-coupling. \cite{Juan:2004,Schurig:2006} 
Moreover, the lattice structure has to be carefully 
chosen as well. In fact, all the double lattice structures listed in 
Table \ref{tab1} have vanishing 
$\mathord{\buildrel{\lower3pt\hbox{$\scriptscriptstyle\leftrightarrow$}}\over 
{\gamma }} _{jj'} $ at ${\rm {\bf q}} = {\rm {\bf 0}}$. From Eq. (\ref{eq5}), Eq. 
(\ref{eq11}) and Eq. (\ref{eq14}), it means that

\begin{equation}
\label{eq15}
\mathord{\buildrel{\lower3pt\hbox{$\scriptscriptstyle\leftrightarrow$}}\over 
{\xi }} \left( {{\rm {\bf q}} = {\rm {\bf 0}}} \right) = 
\mathord{\buildrel{\lower3pt\hbox{$\scriptscriptstyle\leftrightarrow$}}\over 
{\zeta }} \left( {{\rm {\bf q}} = {\rm {\bf 0}}} \right) = 0.
\end{equation}

\noindent
For simplicity, we further assume that the electric particle can only be 
electrically polarized along the x-direction with polarizability $\alpha _e$ 
($c_0 p_{1x}=\alpha _e E_{local,x}(\bf{\tau}_1)/\eta_0$), 
the magnetic particle can only be magnetically polarized along the 
y-direction with polarizability $\alpha _m $ 
($m_{2y}=\alpha _m H_{local,y}(\bf{\tau}_2)$) and all the polarizabilities 
along the other directions are assumed to be negligible. In this case, the 
relevant basis vectors can be reduced to $\left\{ {c_0 p_{1x} ,m_{2y} } 
\right\}$ in Eq. (\ref{eq4}). Eq. (\ref{eq10}) in the local limit can then be written as

\begin{equation}
\label{eq16}
\left\langle {T} \right\rangle ^{ - 1}\left( {{\rm {\bf q}} = {\rm {\bf 0}}} 
\right)\left[ {{\begin{array}{*{20}c}
 {c_0 \left\langle {P_x } \right\rangle } \hfill \\
 {\left\langle {M_y } \right\rangle } \hfill \\
\end{array} }} \right] = \left[ {{\begin{array}{*{20}c}
 {E_{\mbox{ext,x}} / \eta _0 } \hfill \\
 {H_{\mbox{ext,y}} } \hfill \\
\end{array} }} \right],
\end{equation}

\noindent
where

\begin{equation}
\label{eq17}
\begin{array}{l}
\left\langle {T} \right\rangle ^{ - 1}\left( {{\rm {\bf q}} = {\rm {\bf 0}}} \right) + {G}_0 \left( {{\rm {\bf q}} = {\rm {\bf 0}}} \right)= \\
\left[ {{\begin{array}{*{20}c}
 {V\alpha _e^{ - 1} + \frac{ik_0^3 V}{6\pi} - \beta _T } \hfill 
& 0 \hfill \\
 0 \hfill & {V\alpha _m^{ - 1} + \frac{ik_0^3 V}{6\pi} - \beta 
_T } \hfill \\
\end{array} }} \right],
\end{array}
\end{equation}

\noindent
is now written in the $\left\{ {c_0 \left\langle {P_x } \right\rangle 
,\left\langle {M_y } \right\rangle } \right\}$ basis. The macroscopic 
electric polarization $\left\langle {\rm {\bf P}} \right\rangle $ is only 
contributed from the first particle while the macroscopic magnetic 
polarization $\left\langle {\rm {\bf M}} \right\rangle $ is only contributed 
from the second particle. Therefore, the local effective medium is governed 
by

\begin{equation}
\label{eq18}
\begin{split}
 \frac{1}{\epsilon _T - 1} + \beta _T & = V\alpha _e^{ - 1} + ik_0^3 V / 
\left( {6\pi } \right), \\ 
 \frac{1}{\mu _T - 1} + \beta _T & = V\alpha _m^{ - 1} + ik_0^3 V / \left( 
{6\pi } \right). \\ 
 \end{split}
\end{equation}

\noindent
Because of the carefully chosen microstructure, 
the local effective medium has exactly the same form to a 
single lattice of one type of particles. \cite{Belov:2005} This is the 
ideal case that the electric and the magnetic particles cannot ``see'' each 
other. Here, $\beta _T $ is the local limit (${\bf q} = {\bf 0} $) of $\beta _T \left( {\rm {\bf 
q}} \right)$ defined in Eq. (\ref{eq7}). It is a function of the normalized frequency 
$\omega a / \left( {2\pi c_0 } \right)$ for the lattice structure and it can 
be well approximated by a polynomial at small frequencies. 
Table \ref{tab2} gives the polynomial expansion of $\beta _T 
$ for various lattice structures. Note that for cubic lattices, $\beta _T 
\to 1 / 3$ at the long wavelength limit so that Eq. (\ref{eq18}) returns to the 
Clausius-Mossotti equation. \cite{Tretyakov:2003,Mahan:2006}

\begin{table}[htbp]
\begin{tabular}
{|l|l|}
\hline
& 
$\beta _T $ \\
\hline
SC& 
$1 / 3 - 5.97\Omega ^2 + 11.8\Omega ^4$ \\
\hline
FCC& 
$1 / 3 - 2.40\Omega ^2 + 1.72\Omega ^4$ \\
\hline
BCC& 
$1 / 3 - 3.82\Omega ^2 + 4.38\Omega ^4$ \\
\hline
\end{tabular}
\caption{Polynomial expansion of $\beta _T $ for different single lattices. The 
normalized frequency is $\Omega = \omega a / \left( {2\pi c_0 } \right)$ and  
$a$ is the lattice constant.}
\label{tab2}
\end{table}

One popular form of the frequency dispersions of the local effective medium 
is given by

\begin{equation}
\label{eq19}
\begin{split}
 \epsilon _T = 1 - \frac{A\omega _e^2 }{\omega ^2 - \omega _e^2 + i\omega 
\Gamma _e }, \\ 
 \mu _T = 1 - \frac{B\omega ^2}{\omega ^2 - \omega _m^2 + i\omega \Gamma _m 
}. \\ 
 \end{split}
\end{equation}

\noindent
Here, Eq. (\ref{eq19}) is still called the Lorentz-model dispersion although 
the resonance term in permeability is modified so that these frequency dispersion forms 
give us the correct low frequency behavior (including the frequency regime near resonance) 
when there are only one electric resonance and one magnetic resonance. \cite{Boardman:2006}
One point to note here is that for a 
passive particle, we have to satisfy $Im\left( {1 / \alpha _e } \right) \le 
- k_0^3 / \left( {6\pi } \right)$ for the electric polarizability and also a 
similar inequality for the magnetic polarizability. \cite{Tretyakov_book:2003} It guarantees a positive 
imaginary part for both the local permittivity and the local permeability 
according to Eq. (\ref{eq18}), i.e. $\Gamma _e $ or $\Gamma _m $ should be zero for 
a lossless medium or positive for a lossy medium under the dipolar model.

In the following, we would like to specify the local effective medium 
{\{}$\epsilon _T $, $\mu _T ${\}} directly instead of specifying the 
polarizabilities {\{}$\alpha _e $, $\alpha _m ${\}}. As an example, we 
consider the case that $\omega _e / \left( {2\pi } \right) = 8GHz$, $A = 
0.89$, $\omega _m / \left( {2\pi } \right) = 8.5GHz$, $B = 0.128$ and 
$\Gamma _e = \Gamma _m = 0$. The underlying microstructure is assumed to be 
a CsCl structure of lattice constant $a = 5mm$. The polarizabilities can be 
found by Eq. (\ref{eq18}) whenever a microscopic description is necessary. 
Fig. 1(a) and (b) show the local effective 
permittivity and the local permeability. Fig. 1(c) 
shows the crystal being viewed from the direction [001] or [010]. The 
uniaxial electric particle (green color) can be polarized along the [100] 
direction and the uniaxial magnetic particle can be polarized along the 
[010] direction. We will start from this configuration in the next section 
when we consider the non-local effective medium.

\begin{figure}[htbp]
\centerline{\includegraphics[width=3.5in]{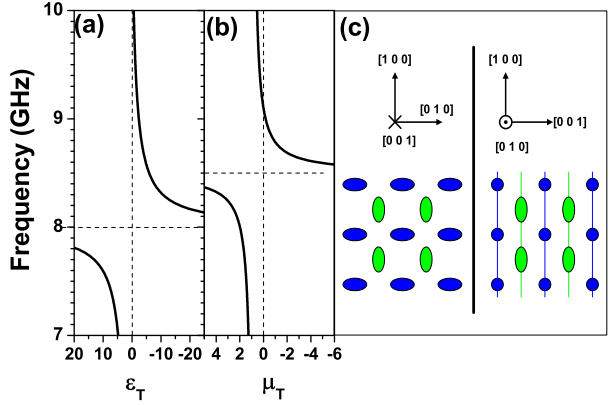}}
\caption{(Color online)A double-lattice configuration specified by Eq. (\ref{eq19}) with $\omega _e / 
\left( {2\pi } \right) = 8GHz$, $A = 0.89$, $\omega _m / \left( {2\pi } 
\right) = 8.5GHz$, $B = 0.128$, $\Gamma _e = \Gamma _m = 0$ and lattice 
constant a= 5 mm. (a) local permittivity; (b) local permeability; (c) The 
underlying CsCl structure viewed along the [001] and the [010] direction. 
The uniaxial electric particle (green color) can be polarized along the 
[100] direction. The uniaxial magnetic particle (blue color) can be 
polarized along the [010] direction. }
\label{fig1}
\end{figure}

It is more convenient to specify the metamaterial by its local effective medium 
directly.
However, a note has to be taken for the freedom of the parameters in the 
frequency dispersion. In particular, for a non-absorptive particle, the 
polarizability can be written in terms of the dipolar scattering phase shift $\delta 
_1^{\left( \sigma \right)} $ (a real number) by

\begin{equation}
\label{eq20}
\frac{i k_{0}^{3} \alpha_{\sigma} }{3\pi } = 
\exp \left( 
  {2i\delta _1^{\left( \sigma \right)} } 
\right) 
- 1,
\end{equation}

\noindent
where $\sigma = e,m$ for ``e'' wave scattering (non-zero local electric field with vanishing local magnetic field at the particle) 
or ``m'' wave scattering (non-zero local magnetic field with vanishing local electric field at the particle). 
Due to the finite volume of one single unit cell, the phase shift $\delta 
_1^{\left( \sigma \right)} $ cannot vary arbitrarily against 
frequency. Suppose a sphere of radius $r_s $ can completely enclose one 
artificial electric particle, the restriction on $\delta _1^{\left( e \right)} $ 
can be written in terms of the time-averaged total energy $\mathscr{E}$ within the 
enclosing sphere (for a quasi-monochromatic electromagnetic field) as:

\begin{equation}
\label{eq21}
\begin{split}
\omega ^3\mathscr{E} \propto & \omega \frac{d\delta _1^{\left( e \right)} }{d\omega } - 
\frac{1}{2\theta ^3}
(1 + 2\theta ^2 - 2\theta ^4 - \cos 2\left( {\theta + \delta _1^{\left( e \right)} } \right) \\
 & - 2\theta\sin 2\left( {\theta + \delta _1^{\left( e \right)} } \right)) \ge 0,
\end{split}
\end{equation}

\noindent
where $\theta = \omega r_s / c_0 $. See Appendix \ref{appendixb} for the derivation. This 
inequality must be satisfied. Otherwise, we can extract energy from the 
artificial particle without first pumping energy to it.

\begin{figure}[htbp]
\centerline{\includegraphics[width=3.5in]{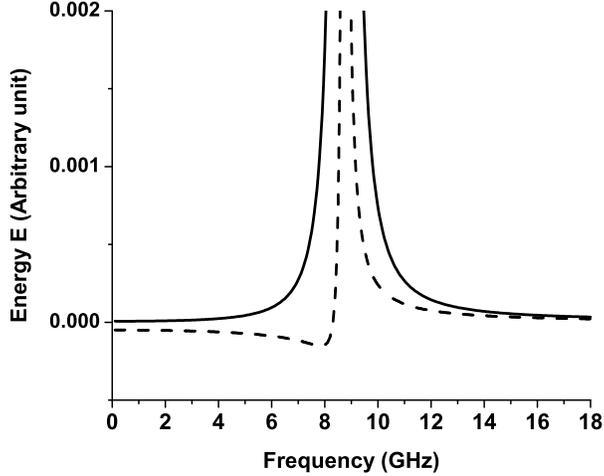}}
\caption{The time-averaged total electromagnetic energy within a sphere of (a)$r_s = 
1.4mm$ and (b) $r_s = 2.0mm$ enclosing the electric particle for ``e'' wave 
scattering.}
\label{fig2}
\end{figure}

For our example, the polarizabilities are obtained through Eq. (\ref{eq18}) 
and the time-averaged total energy for ``e'' wave scattering (in an arbitrary unit) 
is obtained from Eq. (\ref{eq20}) and Eq. (\ref{eq21}). 
Fig. \ref{fig2} shows the results for two different 
cases. The solid line shows the case with $r_s = 2mm$. The total energy 
within the sphere is positive in the whole frequency regime. The dashed line 
shows the case with a smaller radius $r_s = 1.4mm$. In this case, the energy becomes negative 
which is not valid. Therefore, in order to design 
our metamaterial, the electric artificial particle must have a certain minimum size.
On the other hand, if we already know the size of the electric artificial particle, 
it imposes a maximum on the electric resonance strength $A$ (in Eq. (\ref{eq19})) that 
the metamaterial can have.

\section{The non-local effective medium -- double lattice}
\label{sect4}

\noindent
In this section, we investigate how the effective medium changes when the 
wave vector is deviated from the Brillouin zone center. In fact, the tensor 
$\mathord{\buildrel{\lower3pt\hbox{$\scriptscriptstyle\leftrightarrow$}}\over 
{\gamma }} _{12} \left( {\rm {\bf q}} \right)$ does not vanish in general. 
For the double lattice we considered in the last section with ${\rm {\bf 
q}}$ along direction [111] or [100]/[010]/[001], Eq. (\ref{eq10}) can still be 
written in the form of Eq. (\ref{eq16}) where

\begin{equation}
\label{eq22}
\begin{array}{l}
\left( {\left\langle {T} \right\rangle ^{ - 1} + {G}_0 } \right)\left( {\rm 
{\bf q}} \right) = \\
\left[ {{\begin{array}{*{20}c}
 {V\alpha _e^{ - 1} + \frac{ik_0^3 V}{6\pi}  - \beta _T \left( 
{\omega ,{\rm {\bf q}}} \right)} \hfill & {\gamma _{12} \left( {\omega ,{\rm 
{\bf q}}} \right)} \hfill \\
 {\gamma _{12}^\ast \left( {\omega ,{\rm {\bf q}}} \right)} \hfill & 
{V\alpha _m^{ - 1} + \frac{ik_0^3 V}{6\pi} - \beta _T \left( 
{\omega ,{\rm {\bf q}}} \right)} \hfill \\
\end{array} }} \right].
\end{array}
\end{equation}

\noindent
This 2-by-2 matrix is in the $\left\{ {c_0 \left\langle {P_x } \right\rangle ,\left\langle 
{M_y } \right\rangle } \right\}$ basis with $\hat {z}$ being the direction 
along ${\rm {\bf q}}$, $\hat {x}$ being the direction the uniaxial electric 
particle can be polarized and $\hat {y}$ being the direction the uniaxial 
magnetic particle can be polarized. Therefore, by comparing to Eq. (\ref{eq17}), the 
non-local effective medium can be written with respect to the local 
effective medium by

\begin{equation}
\label{eq23}
\begin{split}
 & \left[ {{\begin{array}{*{20}c}
 {\epsilon _T \left( {\rm {\bf q}} \right) - 1} \hfill & { - \xi \left( 
{\rm {\bf q}} \right)} \hfill \\
 { - \zeta \left( {\rm {\bf q}} \right)} \hfill & {\mu _T \left( {\rm {\bf 
q}} \right) - 1} \hfill \\
\end{array} }} \right]^{ - 1} = \\
 & \left[ {{\begin{array}{*{20}c}
 {\frac{1}{\epsilon _T - 1} - \Delta \beta _T \left( {\omega ,{\rm {\bf 
q}}} \right)} \hfill & {\gamma _{12} \left( {\omega ,{\rm {\bf q}}} \right)} 
\hfill \\
 {\gamma _{12}^\ast \left( {\omega ,{\rm {\bf q}}} \right)} \hfill & 
{\frac{1}{\mu _T - 1} - \Delta \beta _T \left( {\omega ,{\rm {\bf q}}} 
\right)} \hfill \\
\end{array} }} \right],
\end{split}
\end{equation}

\noindent
where $\epsilon _T \left( {\rm {\bf q}} \right)$, $\mu _T \left( {\rm {\bf q}} \right)$, 
$\xi \left( {\rm {\bf q}} \right)$ and $\zeta \left( {\rm {\bf q}} \right)$ are the elements 
of the four constitutive tensors in $\left\{ {\hat {x},\hat {y},\hat {z}} \right\}$ basis:

\begin{equation}
\label{eq24}
\begin{array}{l}
 
\mathord{\buildrel{\lower3pt\hbox{$\scriptscriptstyle\leftrightarrow$}}\over 
{\epsilon }} \left( {\rm {\bf q}} \right) = \left[ 
{{\begin{array}{*{20}c}
 {\epsilon _T \left( {\rm {\bf q}} \right)} \hfill & 0 \hfill & 0 \hfill 
\\
 0 \hfill & 1 \hfill & 0 \hfill \\
 0 \hfill & 0 \hfill & 1 \hfill \\
\end{array} }} 
\right];\mathord{\buildrel{\lower3pt\hbox{$\scriptscriptstyle\leftrightarrow$}}\over 
{\xi }} \left( {\rm {\bf q}} \right) = \left[ {{\begin{array}{*{20}c}
 0 \hfill & { - \xi \left( {\rm {\bf q}} \right)} \hfill & 0 \hfill \\
 0 \hfill & 0 \hfill & 0 \hfill \\
 0 \hfill & 0 \hfill & 0 \hfill \\
\end{array} }} \right] \\ 
 
\mathord{\buildrel{\lower3pt\hbox{$\scriptscriptstyle\leftrightarrow$}}\over 
{\zeta }} \left( {\rm {\bf q}} \right) = \left[ {{\begin{array}{*{20}c}
 0 \hfill & 0 \hfill & 0 \hfill \\
 { - \zeta \left( {\rm {\bf q}} \right)} \hfill & 0 \hfill & 0 \hfill \\
 0 \hfill & 0 \hfill & 0 \hfill \\
\end{array} }} 
\right];\mathord{\buildrel{\lower3pt\hbox{$\scriptscriptstyle\leftrightarrow$}}\over 
{\mu }} \left( {\rm {\bf q}} \right) = \left[ {{\begin{array}{*{20}c}
 1 \hfill & 0 \hfill & 0 \hfill \\
 0 \hfill & {\mu _T \left( {\rm {\bf q}} \right)} \hfill & 0 \hfill \\
 0 \hfill & 0 \hfill & 1 \hfill \\
\end{array} }} \right] \\ 
 \end{array}
\end{equation}

\noindent
and $\Delta \beta _T \left( {\omega ,{\rm {\bf q}}} \right)$ is defined by 
$\Delta \beta _T \left( {\omega ,{\rm {\bf q}}} \right) = \beta _T \left( 
{\omega ,{\rm {\bf q}}} \right) - \beta _T \left( {\omega ,{\rm {\bf q}} = 
{\rm {\bf 0}}} \right)$. 
$\mathord{\buildrel{\lower3pt\hbox{$\scriptscriptstyle\leftrightarrow$}}\over 
{\epsilon }} \left( {\rm {\bf q}} \right)$ in the y/z direction and 
$\mathord{\buildrel{\lower3pt\hbox{$\scriptscriptstyle\leftrightarrow$}}\over 
{\mu }} \left( {\rm {\bf q}} \right)$ in the x/z direction have the value 
one since we have neglected the polarizabilities in these directions for 
mathematical simplicity.

\begin{figure}[htbp]
\centerline{\includegraphics[width=3.5in]{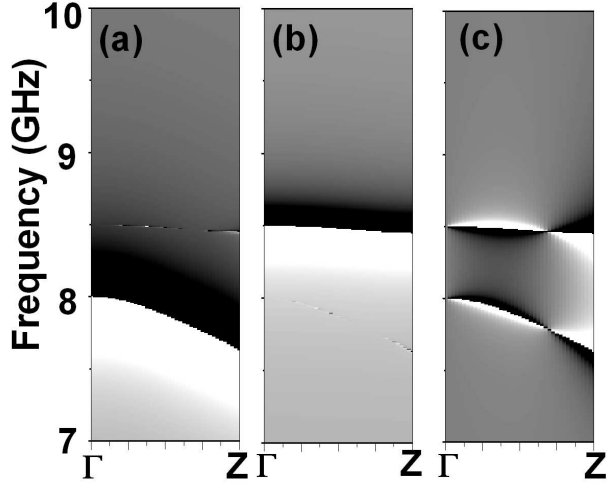}}
\caption{The non-local effective medium from the $\Gamma $ to the Z[001] point for the 
configuration shown in the caption of Fig. 1: (a) 
$\epsilon _T \left( {\rm {\bf q}} \right)$; (b) $\mu _T \left( {\rm {\bf 
q}} \right)$ and (c) $\xi \left( {\rm {\bf q}} \right)$. White color denotes 
positive and black color denotes negative values. }
\label{fig3}
\end{figure}

Now, we use Eq. (\ref{eq23}) to find the non-local effective medium for the previous 
example shown in Fig. 1. In this case, we plot 
$\epsilon _T \left( {\rm {\bf q}} \right)$, $\mu _T \left( {\rm {\bf q}} 
\right)$ and $\xi \left( {\rm {\bf q}} \right)$ along the $\Gamma $ to 
Z[001] direction in Fig. \ref{fig3}. The profile of 
$\epsilon _T \left( {\rm {\bf q}} \right)$ and $\mu _T \left( {\rm {\bf 
q}} \right)$ are shown in Fig. \ref{fig3} (a) and (b). For 
every wave vector ${\rm {\bf q}}$, we can see that $\epsilon _T \left( 
{\rm {\bf q}} \right)$ and $\mu _T \left( {\rm {\bf q}} \right)$ have the 
similar frequency dispersion as the local medium correspondence $\epsilon 
_T $ and $\mu _T $, only with the resonating frequency being shifted. This 
qualitative behavior is expected from Eq. (\ref{eq23}) by neglecting the presence 
of $\gamma _{12} \left( {\rm {\bf q}} \right)$. However, with careful 
examination, we can still see that when ${\rm {\bf q}}$ deviates 
from the $\Gamma $ point, $\epsilon _T \left( {\rm {\bf q}} \right)$ also 
diverges at the magnetic resonating frequency. Moreover, 
we have non-vanishing $\xi \left( {\rm {\bf q}} \right)$ near the electric 
or the magnetic resonating frequency. These result from the presence 
of the term $\gamma _{12} \left( {\rm {\bf q}} \right)$ which means the 
cross-coupling between the electric and the magnetic field due to spatial 
dispersion.

\begin{figure}[htbp]
\centerline{\includegraphics[width=3.5in]{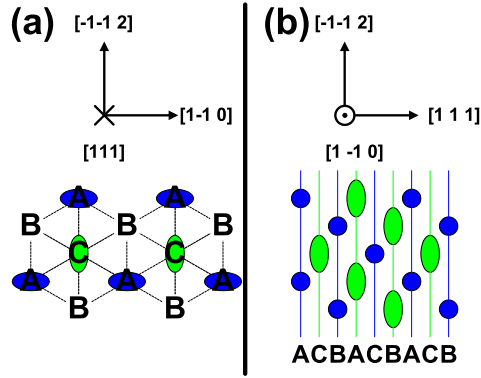}}
\caption{(Color online)A double-lattice configuration specified by Eq. (\ref{eq19}) with $\omega _e / 
\left( {2\pi } \right) = 8GHz$, $A = 0.89$, $\omega _m / \left( {2\pi } 
\right) = 8.5GHz$, $B = 0.128$, $\Gamma _e = \Gamma _m = 0$ and lattice 
constant a= 5 mm. The underlying CsCl structure is viewed along the (a) 
[111] and (b) [0-10] direction. The uniaxial electric particle (green color) 
can be polarized along the [-112] direction. The uniaxial magnetic particle 
(blue color) can be polarized along the [1-10] direction. }
\label{fig4}
\end{figure}

\begin{figure}[htbp]
\centerline{\includegraphics[width=3.5in]{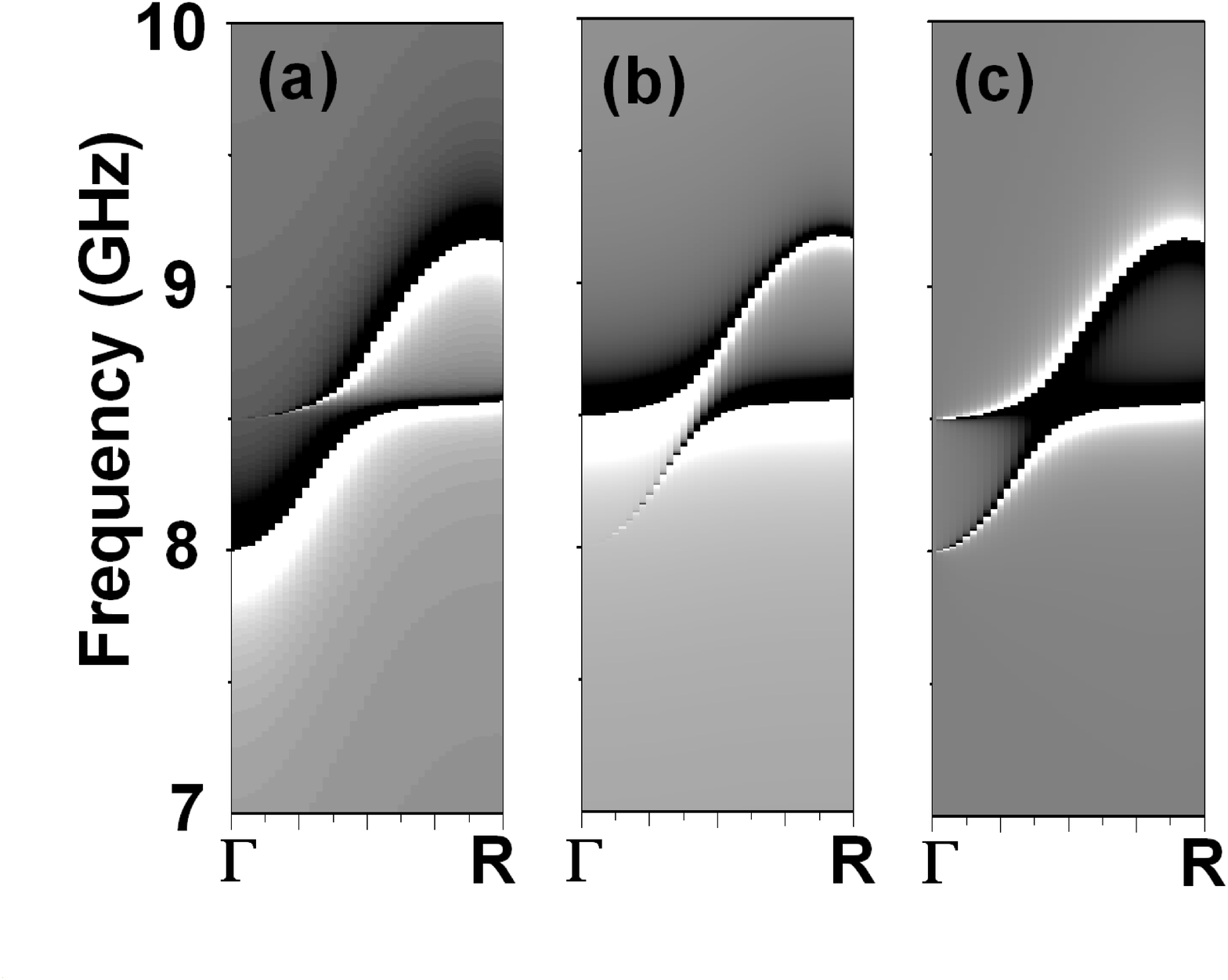}}
\caption{The non-local effective medium from the $\Gamma $ to the R point for the 
configuration shown in the caption of Fig. \ref{fig4}: (a) 
$\epsilon _T \left( {\rm {\bf q}} \right)$; (b) $\mu _T \left( {\rm {\bf 
q}} \right)$ and (c) $\xi \left( {\rm {\bf q}} \right)$. White color denotes 
positive and black color denotes negative values.}
\label{fig5}
\end{figure}

\begin{figure}[htbp]
\centerline{\includegraphics[width=3.5in]{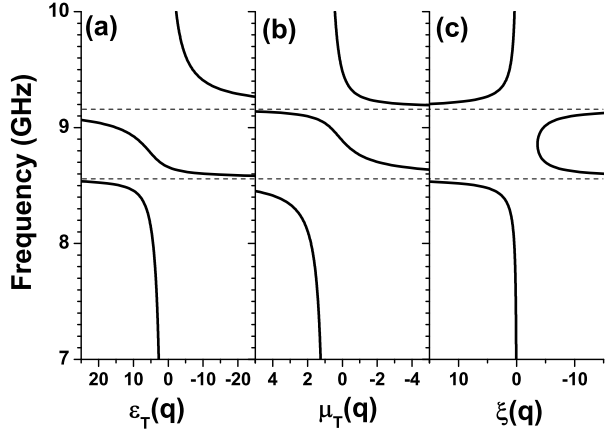}}
\caption{The non-local effective medium at the R point for the configuration shown in 
the caption of Fig. \ref{fig4}: (a) $\epsilon _T \left( 
{\rm {\bf q}} \right)$; (b) $\mu _T \left( {\rm {\bf q}} \right)$ and (c) 
$\xi \left( {\rm {\bf q}} \right)$.}
\label{fig6}
\end{figure}

For the current configuration, the effect of $\gamma _{12} \left( {\rm {\bf 
q}} \right)$ is rather weak. However, it can become more prominent in other 
cases. As an example, for the same CsCl lattice structure, we now consider 
the wavevector ${\rm {\bf q}}$ along the $\Gamma $ to R[111] direction and 
the particles are now orientated so that the uniaxial electric particle can 
be polarized along the [-112] direction while the uniaxial magnetic particle 
can be polarized along the [1-10] direction. This second configuration is 
shown in Fig. \ref{fig4}. $\epsilon _T \left( {\rm {\bf 
q}} \right)$, $\mu _T \left( {\rm {\bf q}} \right)$ along the $\Gamma $ to 
R[111] direction are now plotted in Fig. \ref{fig5} . In 
Fig. \ref{fig6}(a) and (b), we also plot $\epsilon _T 
\left( {\rm {\bf q}} \right)$ and $\mu _T \left( {\rm {\bf q}} \right)$ 
against frequency at exactly the R point. From the results, the resonating 
frequencies depend on ${\rm {\bf q}}$ and both $\epsilon _T \left( {\rm 
{\bf q}} \right)$ and $\mu _T \left( {\rm {\bf q}} \right)$ now clearly 
diverge at two separate frequencies (instead of one) for every non-zero 
${\rm {\bf q}}$ due to the appearance of the term $\gamma _{12} \left( {\rm 
{\bf q}} \right)$. The two resonating frequencies $\omega _1 \left( {\rm 
{\bf q}} \right)$ (the lower one) and $\omega _2 \left( {\rm {\bf q}} 
\right)$ (the upper one) can be obtained from Eq. (\ref{eq23}) and are governed by

\begin{equation}
\label{eq25}
\begin{split}
\left( {\frac{1}{\epsilon _T - 1} - \Delta \beta _T \left( {\rm {\bf q}} 
\right)} \right)\left( {\frac{1}{\mu _T - 1} - \Delta \beta _T \left( {\rm 
{\bf q}} \right)} \right) \\
- \left| {\gamma _{12} \left( {\rm {\bf q}} 
\right)} \right|^2 = 0.
\end{split}
\end{equation}

\noindent
If the magnetic resonance and the electric resonance are far apart in 
frequency or the term $\gamma _{12} \left( {\rm {\bf q}} \right)$ can be 
neglected, it decouples to

\begin{equation}
\label{eq26}
\frac{1}{\epsilon _T - 1} - \Delta \beta _T \left( {\rm {\bf q}} \right) 
= 0,
\end{equation}

\noindent
which determines the electric resonating frequency and

\begin{equation}
\label{eq27}
\frac{1}{\mu _T - 1} - \Delta \beta _T \left( {\rm {\bf q}} \right) = 0,
\end{equation}

\noindent
which determines the magnetic resonating frequency. It gives us two separate 
``bands'' of resonating frequencies. The term $\Delta \beta _T \left( {\rm 
{\bf q}} \right)$ gives us the ${\rm {\bf q}}$ dependence of the electric or 
the magnetic resonating frequency (shown in the previous example in 
Fig. \ref{fig3}). However, in the current example, these two 
bands hybridize with each other so that $\epsilon _T \left( {\rm {\bf q}} 
\right)$ or $\mu _T \left( {\rm {\bf q}} \right)$ has two resonating 
frequencies instead of one. It is called the cross-coupled resonance here.

Due to the same reason, the $\gamma _{12} \left( {\rm {\bf q}} \right)$ term 
also causes a structure-induced bianisotropy when we bring the electric and 
magnetic resonances near to each other in frequency. 
This bianisotropy is induced by the structure instead of the cross-coupling 
effect of the artificial atoms. 
The $\xi \left( {\rm {\bf q}} \right)$ from the $\Gamma $ to the R point is plotted in 
Fig. \ref{fig5}(c). It also diverges at $\omega _1 \left( 
{\rm {\bf q}} \right)$ and $\omega _2 \left( {\rm {\bf q}} \right)$.

We have used a Lorentz-model dispersion (Eq. (\ref{eq19})) for the local 
effective medium. In fact, $\epsilon _T \left( {\rm {\bf q}} \right)$, 
$\mu _T \left( {\rm {\bf q}} \right)$ and $\xi \left( {\rm {\bf q}} \right)$ 
can be numerically fitted very well by extending the same model:

\begin{equation}
\label{eq28}
\epsilon _T \left( {\rm {\bf q}} \right) \approx 1 - 
\sum\limits_{j = 1}^2 
\frac{A_j \left( {\rm {\bf q}} \right)\omega _j^2 \left( {\rm {\bf q}} \right)}{\omega ^2 - 
\omega _j^2 \left( {\rm {\bf q}} \right) + i\omega \Gamma _j \left( {\rm 
{\bf q}} \right)},
\end{equation}

\begin{equation}
\label{eq29}
\mu _T \left( {\rm {\bf q}} \right) \approx 1 - 
\sum\limits_{j = 1}^2 
\frac{B_j \left( {\rm {\bf q}} \right)\omega ^2}{\omega ^2 - \omega _j^2 \left( {\rm {\bf q}} \right) + 
i\omega \Gamma _j \left( {\rm {\bf q}} \right)},
\end{equation}

\noindent
and

\begin{equation}
\label{eq30}
\xi \left( {\rm {\bf q}} \right) \approx C\left( {\rm {\bf q}} \right)\omega ^3
\prod\limits_{j = 1}^2 
\frac{1}{\omega ^2 - \omega _j^2 \left( {\rm {\bf q}} \right) + i\omega \Gamma _j \left( {\rm {\bf q}} \right)}.
\end{equation}

\noindent
In the current example at the $R$ point, the expressions obtained by setting 
$\omega _1 \left( {\rm {\bf q}} \right) / \left( {2\pi } \right) = 8.57GHz$, 
$\omega _2 \left( {\rm {\bf q}} \right) / \left( {2\pi } \right) = 9.17GHz$, 
$A_1 \left( {\rm {\bf q}} \right) = 0.13$, $A_2 \left( {\rm {\bf q}} \right) 
= 0.55$, $B_1 \left( {\rm {\bf q}} \right) = 0.10$, $B_2 \left( {\rm {\bf 
q}} \right) = 0.03$, $C\left( {\rm {\bf q}} \right)2\pi = 0.15GHz^{ - 
1}$ and $\Gamma _1 \left( {\rm {\bf q}} \right) = \Gamma _2 \left( {\rm {\bf 
q}} \right) = 0$ can fit the actual results very well. The fitted result has 
no noticeable differences from the results shown in 
Fig. \ref{fig6}.

\begin{figure}[htbp]
\centerline{\includegraphics[width=3.5in]{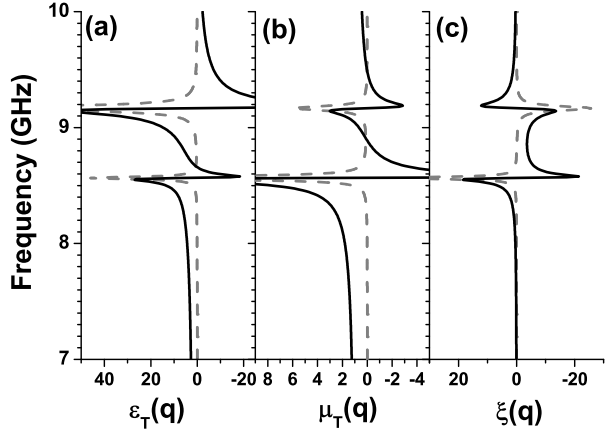}}
\caption{(Color online)The non-local effective medium at the R point for the configuration shown in 
the caption of Fig. \ref{fig4} with $\Gamma _e = 0.05$ and 
$\Gamma _m = 0.02$: (a) $\epsilon _T \left( {\rm {\bf q}} \right)$ ; (b) 
$\mu _T \left( {\rm {\bf q}} \right)$; (c) $\xi \left( {\rm {\bf q}} 
\right)$. The blue solid line shows the real part and the gray dashed line 
shows the imaginary part.}
\label{fig7}
\end{figure}

The extended Lorentz-model dispersion is valid even for an absorptive system. Suppose, now we 
add some absorption to the system by setting $\Gamma _e / \left( {2\pi } 
\right) = 0.05GHz$ and $\Gamma _m / \left( {2\pi } \right) = 0.02GHz$ in Eq. (\ref{eq19}), the 
resultant non-local effective medium at the $R$ point is shown in 
Fig. \ref{fig7} . The non-local effective medium can still 
be fitted by the extended Lorentz model but with non-zero 
$\Gamma _1 \left( {\rm {\bf q}} \right)$ and $\Gamma _2 \left( {\rm {\bf q}} 
\right)$. In this case, we set $\Gamma _1 \left( {\rm {\bf q}} \right) / 
\left( {2\pi } \right) = 0.025GHz$ and $\Gamma _2 \left( {\rm {\bf q}} 
\right) / \left( {2\pi } \right) = 0.045GHz$ while all other parameters 
remain the same. The fitted expressions also show no noticeable differences 
to the results shown in Fig. \ref{fig7} . 

From Fig. \ref{fig7}, we see that the imaginary part of both 
$\epsilon _T \left( {\rm {\bf q}} \right)$ and $\mu _T \left( {\rm {\bf q}} \right)$ are positive.
For a bianisotropic medium, there are no general requirements that the imaginary part of 
the permittivity and the permeability should be positive. 
However, within the dipolar model, from Eq. 
(\ref{eq23}) with $\beta _T \left( {\omega ,{\rm {\bf q}}} \right)$ being real, it is easy to show that both $\epsilon _T \left( {\rm {\bf q}} \right)$ 
and $\mu _T \left( {\rm {\bf q}} \right)$ have positive imaginary part if 
the local $\epsilon _T $ and $\mu _T $ have positive imaginary part, i.e. 
$\Gamma _1 \left( {\rm {\bf q}} \right)$ and $\Gamma _2 \left( {\rm {\bf q}} 
\right)$ should be positive. In our approach, we parameterize the effective 
medium by using ${\rm {\bf q}}$ such that ${\rm {\bf q}}$ is always a real vector within 
the first Brillouin zone. If, on the other hand, ${\rm {\bf q}}$ is allowed to be complex 
(e.g. traveling along the dispersion of an eigenmode and becoming complex within the 
resonance gap where no propagating eigenmodes can be found), a negative imaginary part in 
$\epsilon _T \left( {\rm {\bf q}} \right)$ or $\mu _T \left( {\rm {\bf 
q}} \right)$ may appear around a cross-coupled resonance. \cite{Koschny:2005}

The structure-induced bianisotropy $\xi \left( {\rm {\bf q}} \right)$ 
appears even if the local effective medium does not suffer bianisotropy. 
$\xi \left( {\rm {\bf q}} \right)$ is generally present and it is purely 
real for a non-absorptive system in our example. It disappears in the local effective medium 
limit only if the particles do not suffer cross-coupling and the particles 
are placed in a carefully chosen lattice. For example, if the sublattices of 
the two types of particles are displaced in the z direction (along ${\rm 
{\bf q}})$ with respect to each other, the 2D symmetry perpendicular to ${\rm {\bf q}}$ does not change 
so that the constitutive tensors can still be described by Eq. (\ref{eq24}). 
However, $\xi \left( {{\rm {\bf q}} = {\rm {\bf 0}}} \right)$ will not 
vanish and it is a purely imaginary number. On the other hand, if the 
lattice is not displaced but the particles suffer cross coupling, $\xi 
\left( {{\rm {\bf q}} = {\rm {\bf 0}}} \right)$ also does not vanish and it 
is purely imaginary. \cite{Xudong:2005}

\begin{figure}[htbp]
\centerline{\includegraphics[width=3.5in]{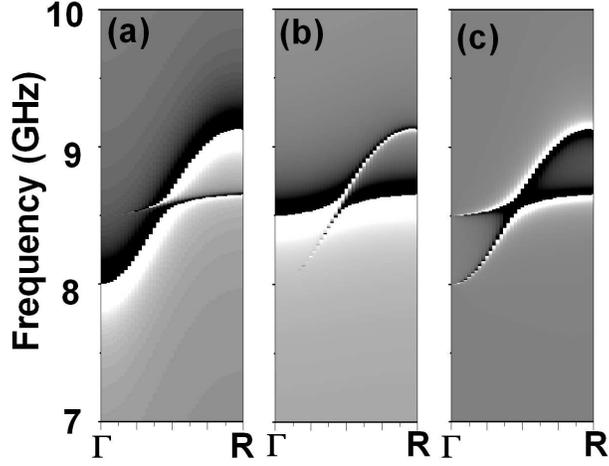}}
\caption{The non-local effective medium from the $\Gamma $ to the R point for the 
configuration shown in the caption of Fig. \ref{fig4} with 
lattice constant a changed to 2.5 mm: (a) $\epsilon _T \left( {\rm {\bf 
q}} \right)$; (b) $\mu _T \left( {\rm {\bf q}} \right)$ and (c) $\xi \left( 
{\rm {\bf q}} \right)$. White color denotes positive and black color denotes 
negative values.}
\label{fig8}
\end{figure}

The structure-induced bianisotropy and cross-coupled resonance are both due 
to the term $\gamma _{12} \left( {\rm {\bf q}} \right)$. In fact, with a
smaller lattice constant, the interaction becomes smaller. 
Fig. \ref{fig8} shows the $\epsilon _T \left( {\rm {\bf 
q}} \right)$, $\mu _T \left( {\rm {\bf q}} \right)$ and $\xi \left( {\rm 
{\bf q}} \right)$ along the $\Gamma $ to R[111] direction when the lattice 
constant $a$ is changed a smaller value 2.5 mm. The splitting between the two 
resonating frequencies is smaller than in Fig. \ref{fig5}. 
On the other hand, if the lattice constant becomes larger, the term $\gamma 
_{12} \left( {\rm {\bf q}} \right)$ cannot be neglected in finding the 
effective medium or in finding the dispersion.

\begin{figure}[htbp]
\centerline{\includegraphics[width=3.5in]{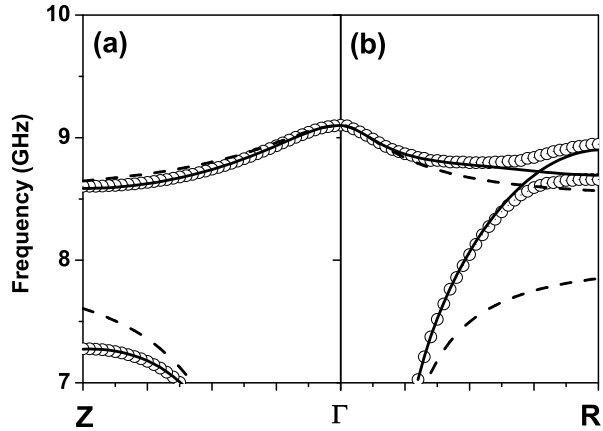}}
\caption{Dispersion diagram (open circles) for the configuration shown in (a) 
Fig. 1 along the $\Gamma Z$ direction and the 
configuration shown in (b) Fig. \ref{fig4} along the $\Gamma 
R$ direction. The black dashed line represents the corresponding dispersion 
obtained from the local effective medium. The blue solid line represents the 
perturbed dispersion from the local one.}
\label{fig9}
\end{figure}

We have investigated the constitutive tensors of the metamaterial. After 
finding the effective medium, the dispersion diagram can also be found. The 
dispersion relation of the effective medium is governed by

\begin{equation}
\label{eq31}
\det \left( {\left\langle {T} \right\rangle ^{ - 1}\left( {\rm {\bf q}} 
\right)} \right) = 0.
\end{equation}

\noindent
After some algebra, we have

\begin{equation}
\label{eq32}
\begin{split}
\det \left( {\left\langle {{T}^{ - 1}} \right\rangle \left( {\rm {\bf q}} 
\right)} \right) & = \frac{1}{q^2 - k_0^2 } \\
 & \times \frac{\left| {q + k_0 \xi \left( 
{\rm {\bf q}} \right)} \right|^2 - k_0^2 \epsilon _T \left( {\rm {\bf q}} 
\right)\mu _T \left( {\rm {\bf q}} \right)}{\left( {\epsilon _T \left( 
{\rm {\bf q}} \right) - 1} \right)\left( {\mu _T \left( {\rm {\bf q}} 
\right) - 1} \right) - \left| {\xi \left( {\rm {\bf q}} \right)} \right|^2}
\end{split}
\end{equation}

\noindent
and the dispersion relation is therefore

\begin{equation}
\label{eq33}
\left| {q + k_0 \xi \left( {\rm {\bf q}} \right)} \right|^2 = k_0^2 
\epsilon _T \left( {\rm {\bf q}} \right)\mu _T \left( {\rm {\bf q}} 
\right).
\end{equation}

\noindent
There is an extra term $k_0 \xi \left( {\rm {\bf q}} \right)$ in the 
dispersion relation comparing to the one obtained from the local effective 
medium: $q^2 = k_0^2 \epsilon _T \mu _T .$ Unlike the case of the local effective 
medium which serves as an approximation of the actual band structure, 
the dispersion relation shown in Eq. (\ref{eq33}) 
can be used to find the band structure which shows no noticeable differences to 
the actual band structure if we solve it directly without first finding the non-local 
effective medium. For band structure of a single lattice consisting artificial particles 
of only electric or only magnetic response, see Ref. \onlinecite{Belov:2005} and Ref. 
\onlinecite{Kempa:2005}.

The dispersion diagram along the direction $\Gamma Z$ for the first 
configuration (Fig. 1) and the dispersion diagram 
along the direction $\Gamma R$ for the second configuration 
(Fig. \ref{fig4}) are shown in Fig. \ref{fig9}(a) and (b) respectively. They are plotted in open black circles while the 
dispersion diagram obtained from the local effective medium is plotted in 
dashed line. Along the $\Gamma Z$ direction for the first 
configuration, they look very similar especially near the Brillouin zone 
center. In fact, the formation of the non-local bands can be understood from 
a perturbation point of view by assuming small $\Delta \beta _T \left( {\rm 
{\bf q}} \right)$ and $\gamma _{12} \left( {\rm {\bf q}} \right)$. For a 
fixed ${\rm {\bf q}}$, we can approximate the change in square frequency 
from the local to the non-local dispersion by

\begin{equation}
\label{eq34}
\begin{split}
\frac{\partial k_0^2 \epsilon _T \mu _T }{\partial k_0^2 }\Delta k_0^2 & = 
\left( {\epsilon _T - 1} \right)\left( {\mu _T - 1} \right)\left( {q^2 - 
k_0^2 } \right) \\ 
 & \times \det \left( {\left\langle {T} \right\rangle ^{ - 1}\left( 
{\rm {\bf q}} \right)} \right),
\end{split}
\end{equation}

\noindent
which is evaluated along one of the bands obtained from the local effective 
medium. (The proof is not shown here.) By substituting Eq. (\ref{eq23}) into 
it and neglect the second order effect of $\beta _T \left( {\rm {\bf 
q}} \right)$ and $\gamma \left( {\rm {\bf q}} \right)$, we have

\begin{equation}
\label{eq35}
\begin{split}
\frac{d\omega ^2\epsilon _T \mu _T }{d\omega ^2}\frac{\Delta \omega 
^2}{\omega ^2} & \approx - \beta _T \left( {\rm {\bf q}} \right)\left( {\mu _T 
\left( {\epsilon _T - 1} \right)^2 + \epsilon _T \left( {\mu _T - 1} 
\right)^2} \right) \\ 
 & + Re\left( {\gamma _{12} \left( {\rm {\bf q}} \right)} 
\right)2\sqrt {\epsilon _T \mu _T } \left( {\epsilon _T - 1} 
\right)\left( {\mu _T - 1} \right).
\end{split}
\end{equation}

\noindent
This perturbed dispersion is plotted in Fig. \ref{fig9} in 
solid line. We can see that it approximates the actual non-local dispersion 
very well in the $\Gamma $Z direction of the first configuration. In the 
$\Gamma $R direction, the first band (having positive refractive index in 
the local limit) and the second band (having negative refractive index in 
the local limit) hybridize with each other to form the actual band.

For every mode in the dispersion diagram, we can characterize it by defining the refractive index 
$n$ of the mode using $q^2 = n^2k_0^2 $ and defining its characteristic impedance Z to be the ratio 
between the tangential $E$ and the tangential $H$ field. They are governed by

\begin{equation}
\label{eq36}
\begin{split}
 n / Z & = \frac{\epsilon _T \left( {\rm {\bf q}} \right)}{1 + \xi \left( 
{\rm {\bf q}} \right)k_0 / q}, \\ 
 nZ & = \frac{\mu _T \left( {\rm {\bf q}} \right)}{1 + \xi \left( {\rm {\bf 
q}} \right)k_0 / q}. \\ 
 \end{split}
\end{equation}

\begin{figure}[htbp]
\centerline{\includegraphics[width=3.5in]{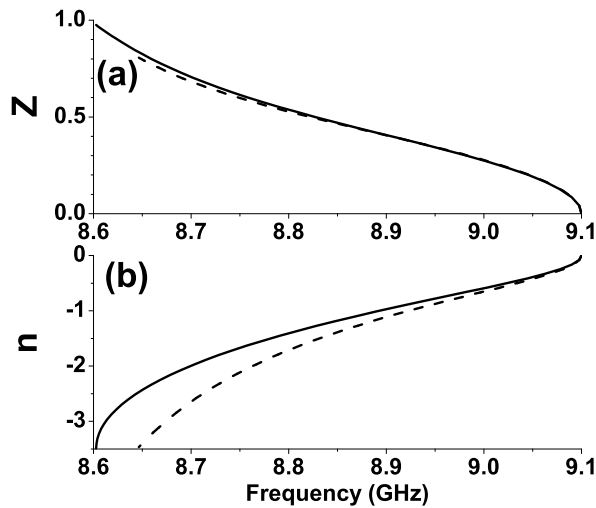}}
\caption{Solid line: characteristic impedance and the refractive index of the double-negative 
band in the $\Gamma Z$ direction. Dashed line: corresponding values for the 
local medium.}
\label{fig10}
\end{figure}

\noindent
For example, the $n$ and $Z$ along the band in the $\Gamma $Z direction in Fig. \ref{fig9} 
are plotted as solid lines in Fig. \ref{fig10} for the double-negative band. The 
corresponding values for the local effective medium ($n^2 = \epsilon _T 
\mu _T $,$Z = \sqrt {\mu _T / \epsilon _T } )$ are also shown in dashed 
lines in the same figure. We can see that the local effective medium works 
very well for the propagating bands unless it is near the Brillouin zone 
edge and near the resonance gap.

We have examined the non-local effective medium governing the transverse waves. 
In fact, if the artificial particles are oriented so that they can only 
be polarized in a single direction along the z-axis, the metamaterial supports longitudinal waves 
whose propagating direction is also along the z-axis. 
We can do the similar analysis to obtain the longitudinal permittivity $\epsilon_L({\bf q})$ 
and the longitudinal permeability $\mu_L({\bf q})$. 
In this case, the electric and magnetic response are decoupled, 
$\epsilon_L({\bf q})$ and $\mu_L({\bf q})$ can also be fitted by using a Lorentz-model dispersion:

\begin{equation}
\label{eq37}
\epsilon _L \left( {\rm {\bf q}} \right) \approx 1 - 
\frac{A_e \left( {\rm {\bf q}} \right)\omega _e^2 \left( {\rm {\bf q}} \right)}{\omega ^2 - 
\omega _e^2 \left( {\rm {\bf q}} \right) + i\omega \Gamma _e \left( {\rm 
{\bf q}} \right)},
\end{equation}

\noindent
and

\begin{equation}
\label{eq38}
\mu _L \left( {\rm {\bf q}} \right) \approx 1 - 
\frac{B_m \left( {\rm {\bf q}} \right)\omega ^2}{\omega ^2 - \omega _m^2 \left( {\rm {\bf q}} \right) + 
i\omega \Gamma _m \left( {\rm {\bf q}} \right)},
\end{equation}

\noindent
The electric longitudinal mode (or the electric bulk plasmon with macroscopic E field along ${\bf q}$) has a dispersion 
relation governed by 
\begin{equation}
\label{eq39}
\epsilon _L(\bf{q}) = 0,
\end{equation}

\noindent
and the dispersion relation for the the magnetic longitudinal mode (or the magnetic bulk plasmon with macroscopic H field along ${\bf q}$) is governed by 
\begin{equation}
\label{eq40}
\mu _L(\bf{q}) = 0.
\end{equation}

\noindent
With spatial dispersion such that $\epsilon _L(\bf{q})$ and 
$\mu _L(\bf{q})$ depends on ${\bf q}$, the band for either the electric or magnetic longitudinal mode becomes 
dispersive to form a narrow band instead of a flat line in the local medium picture.

\section{Conclusion}
\label{sect5}

\noindent
We have established a numerical method based on a dipolar model to obtain the non-local effective 
medium of a metamaterial. In particular, we have concentrated on the double lattice structure in which 
one sublattice holds the electric artificial atoms and another sublattice holds the magnetic artificial atoms.
Once the dipolar scattering properties of the atoms or the local effective medium parameters are specified, the 
non-local effective medium with all the four constitutive tensors can be obtained on the whole frequency regime 
including frequencies near the resonance gap.
We found that the metamaterial should be regarded as bianisotropic near the resonance gap even if the artificial 
atoms do not suffer cross-coupling. 
In this case, the cross-coupling comes from spatial dispersion and it also induces a cross-coupled resonance for metamaterials 
having both electric and magnetic resonances. Within our model, the cross-coupled resonance can still be understood 
using a Lorentz dispersion model. The effect of the cross-coupling from spatial dispersion depends on the microstructure and 
it is larger at a higher normalized frequency (or a larger lattice constant).

\section{Acknowledgment}

This work is supported by the Croucher Foundation fellowship from Hong Kong.

\appendix
\section{The Green's Tensor}
\label{appendixa}

The radiation field from an electric dipole ${\rm {\bf p}}$ together with a 
magnetic dipole ${\rm {\bf m}}$ at the origin in the vacuum is

\begin{equation}
\label{eqa1}
\left[ {{\begin{array}{*{20}c}
 {{\rm {\bf E}}\left( {\rm {\bf r}} \right) / \eta _0 } \hfill \\
 {{\rm {\bf H}}\left( {\rm {\bf r}} \right)} \hfill \\
\end{array} }} \right] = {G}_0 \left( {\rm {\bf r}} \right) \cdot \left[
{{\begin{array}{*{20}c}
 {c_0{\rm {\bf p}}} \hfill \\
 {\rm {\bf m}} \hfill \\
\end{array} }} \right],
\end{equation}

\noindent
where the ${G}_0 $ matrix is defined by

\begin{equation}
\label{eqa2}
{G}_0 \left( {\rm {\bf r}} \right) = \left[ {{\begin{array}{*{20}c}
 {k_0^2 
\mathord{\buildrel{\lower3pt\hbox{$\scriptscriptstyle\leftrightarrow$}}\over 
{G}} _0 \left( {\rm {\bf r}} \right)} \hfill & {ik_0 \nabla \times 
\mathord{\buildrel{\lower3pt\hbox{$\scriptscriptstyle\leftrightarrow$}}\over 
{G}} _0 \left( {\rm {\bf r}} \right)} \hfill \\
 { - ik_0 \nabla \times 
\mathord{\buildrel{\lower3pt\hbox{$\scriptscriptstyle\leftrightarrow$}}\over 
{G}} _0 \left( {\rm {\bf r}} \right)} \hfill & {k_0^2 
\mathord{\buildrel{\lower3pt\hbox{$\scriptscriptstyle\leftrightarrow$}}\over 
{G}} _0 \left( {\rm {\bf r}} \right)} \hfill \\
\end{array} }} \right],
\end{equation}

\noindent
with the vacuum Green's tensor 
$\mathord{\buildrel{\lower3pt\hbox{$\scriptscriptstyle\leftrightarrow$}}\over 
{G}} _0 $ defined by

\begin{equation}
\label{eqa3}
\mathord{\buildrel{\lower3pt\hbox{$\scriptscriptstyle\leftrightarrow$}}\over 
{G}} _0 \left( {\rm {\bf r}} \right) = \left( 
{\mathord{\buildrel{\lower3pt\hbox{$\scriptscriptstyle\leftrightarrow$}}\over 
{I}} + \frac{1}{k_0^2 }\nabla \nabla } \right)\frac{\exp \left( {ik_0 r} 
\right)}{4\pi r},
\end{equation}

\noindent
where 
$\mathord{\buildrel{\lower3pt\hbox{$\scriptscriptstyle\leftrightarrow$}}\over 
{I}} $ is the 3-by-3 identity tensor.

\noindent
The Fourier Transform of ${G}_0 \left( {\rm {\bf r}} \right)$ is given by

\begin{equation}
\label{eqa4}
\begin{split}
 {G}_0 \left( {\rm {\bf q}} \right) & = \int {{G}_0 \left( {\rm {\bf r}} 
\right)\exp \left( { - i{\rm {\bf q}} \cdot {\rm {\bf r}}} \right)d^3r} \\ 
 & = \left[ {{\begin{array}{*{20}c}
 {k_0^2 
\mathord{\buildrel{\lower3pt\hbox{$\scriptscriptstyle\leftrightarrow$}}\over 
{G}} _0 \left( {\rm {\bf q}} \right)} \hfill & { - k_0 {\rm {\bf q}}\times 
\mathord{\buildrel{\lower3pt\hbox{$\scriptscriptstyle\leftrightarrow$}}\over 
{G}} _0 \left( {\rm {\bf q}} \right)} \hfill \\
 {k_0 {\rm {\bf q}}\times 
\mathord{\buildrel{\lower3pt\hbox{$\scriptscriptstyle\leftrightarrow$}}\over 
{G}} _0 \left( {\rm {\bf q}} \right)} \hfill & {k_0^2 
\mathord{\buildrel{\lower3pt\hbox{$\scriptscriptstyle\leftrightarrow$}}\over 
{G}} _0 \left( {\rm {\bf q}} \right)} \hfill \\
\end{array} }} \right], \\ 
\end{split}
\end{equation}

\noindent
where

\begin{equation}
\label{eqa5}
\mathord{\buildrel{\lower3pt\hbox{$\scriptscriptstyle\leftrightarrow$}}\over 
{G}} _0 \left( {\rm {\bf q}} \right) = \frac{1}{q^2 - k_0^2 }\left( 
{\mathord{\buildrel{\lower3pt\hbox{$\scriptscriptstyle\leftrightarrow$}}\over 
{I}} - \hat {q}\hat {q}} \right) - \frac{1}{k_0^2 }\hat {q}\hat {q}.
\end{equation}

\section{Restriction on the scattering phase shift of a single particle}
\label{appendixb}

Here, we will consider the restriction on the polarizability for a lossless 
classical uniaxial particle which can be described by its permittivity distribution 
$\epsilon _s \left( {\rm {\bf r}} \right)$ and its permeability 
distribution $\mu _s \left( {\rm {\bf r}} \right)$. The particle has a finite radius 
$r_s$ such that it is vacuum outside the sphere of radius $r_s$ centered at the origin.
Without losing generality, we consider the situation that ``m'' wave (with angular momentum 
$l = 1$ and $m = 0)$ is scattered from the particle which can be 
magnetically polarized along the z-direction but not the other directions. 
We also assume that the particle has no cross-coupling, i.e. the electric 
dipole moment is zero when the local electric field is zero here.
The total wave outside the particle can be written in the following form:

\begin{equation}
\label{eqb1}
{\rm {\bf E}} = \frac{u\left( k_0 r \right)}{r}{\rm {\bf r}}\times \nabla 
Y_{l = 1,m = 0} \left( \hat {r} \right)
\end{equation}

\noindent
where

\begin{equation}
\label{eqb2}
u\left( \theta \right) = \frac{\theta }{2}\left( {h_1^\ast \left( \theta 
\right) + \exp \left( {2i\delta _1^{\left( m \right)} } \right)h_1 \left( 
\theta \right)} \right),
\end{equation}

\noindent
with $h_l \left( \theta \right)$ being the out-going spherical Hankel 
function, and $Y_{lm} \left( \hat {r} \right)$ being the spherical harmonic. 
$\delta _l^{\left( m \right)} $,  the phase shift in scattering, is a real number for a 
lossless particle and it is related to the magnetic polarizability through

\begin{equation}
\label{eqb3}
\frac{i k_{0}^{3} \alpha_{m} }{3\pi } = 
\exp \left( 
  {2i\delta _1^{\left( m \right)} } 
\right) 
- 1.
\end{equation}

\noindent
On the other hand, from the Maxwell equations, we can prove the following 
identity

\begin{equation}
\label{eqb4}
\begin{split}
\frac{1}{4}\int {\epsilon _0 \frac{\partial \omega \epsilon \left( 
{\rm {\bf r}} \right)}{\partial \omega }\left| {\rm {\bf E}} \right|^2 + \mu 
_0 \frac{\partial \omega \mu \left( {\rm {\bf r}} \right)}{\partial \omega 
}\left| {\rm {\bf H}} \right|^2d^3r }= \\
- \frac{i}{4}\oint {\left( 
{\frac{\partial {\rm {\bf E}}}{\partial \omega }\times {\rm {\bf H}}^\ast + 
{\rm {\bf E}}^\ast \times \frac{\partial {\rm {\bf H}}}{\partial \omega }} 
\right) \cdot d{\rm {\bf A}}}  .
\end{split}
\end{equation}

\noindent
The volume and the surface integrals are taken to be the volume and the 
surface of the sphere of radius $r_s $ which encloses the particle completely. 
By substituting Eq. (\ref{eqb1}) into Eq. (\ref{eqb4}), we obtain

\begin{equation}
\begin{split}
\label{eqb5}
\frac{1}{4}\int {\epsilon _0 \frac{\partial \omega \epsilon \left( 
{\rm {\bf r}} \right)}{\partial \omega }\left| {\rm {\bf E}} \right|^2 + \mu 
_0 \frac{\partial \omega \mu \left( {\rm {\bf r}} \right)}{\partial \omega 
}\left| {\rm {\bf H}} \right|^2d^3r} = \\
\frac{1}{4\omega ^2\mu _0 }\left( 
{{u}'^\ast \left( \theta \right)\frac{\partial u\left( \theta 
\right)}{\partial \omega } - \frac{\partial {u}'\left( \theta 
\right)}{\partial \omega }u^\ast \left( \theta \right)} \right), 
\end{split}
\end{equation}

\noindent
where $\theta = \omega r_s / c$. Finally, we recognize the l.h.s. of Eq. (\ref{eqb5}) is just the time-averaged total 
electromagnetic energy ($\mathscr{E})$ within the sphere for a narrow band signal 
centered at the angular frequency $\omega$. 
By substituting Eq. (\ref{eqb2}) into Eq. (\ref{eqb5}), we can express the time-averaged total energy 
in terms of $\delta _l^{\left( m \right)} $:

\begin{equation}
\label{eqb6}
\begin{split}
\omega ^3\mathscr{E} \propto & \omega \frac{d\delta _1^{\left( m \right)} }{d\omega } - 
\frac{1}{2\theta ^3} (1 + 2\theta ^2 - 2\theta ^4 - \cos 2\left( {\theta + \delta _1^{\left(m \right)} } \right) \\
 & - 2\theta\sin 2\left( {\theta + \delta _1^{\left( m \right)}}\right) ),
\end{split}
\end{equation}

\noindent
which must be a positive number. The derivation for the electric scattering 
phase shift is similar. Moreover, the derivation also applies for an isotropic particle 
instead of an uniaxial particle being discussed here.


\begin{thebibliography}{2}
\bibitem{Smith:2005} D. R. Smith, D. C. Vier, Th. Koschny, and C. M. Soukoulis, \textit{Phys. Rev. E} \textbf{71}, 036617 (2005).
\bibitem{Pendry:1999} J. B. Pendry, A. J. Holden, D. J. Robbins, and W. J. Stewart, \textit{IEEE Trans. Microwave Theory Tech.} \textbf{47}, 2075 (1999).
\bibitem{Smith:2006} D. R. Smith, J. Gollub, J. J. Mock, W. J. Padilla and D. Schurig , \textit{J. Appl. Phys.} \textbf{100}, 024507 (2006).
\bibitem{Lerat:2006} J.-M. Lerat, N. Mall\'{e}jac, and O. Acher, \textit{J. Appl. Phys.} \textbf{100}, 084908 (2006).
\bibitem{Belov:2006} P. A. Belov and C. R. Simovski, \textit{Phys. Rev. B} \textbf{73}, 045102 (2006).
\bibitem{Belov:2003} P. A. Belov, R. Marques, S. I. Maslovski, I. S. Nefedov, M. Silveirinha, C. R. Simovski, and S. A. Tretyakov, \textit{Phys. Rev. B} \textbf{67} 113103 (2003).
\bibitem{Silveirinha:2006} M. G. Silveirinha, \textit{Phys. Rev. E} \textbf{73}, 046612 (2006).
\bibitem{Shapiro:2006} M. A. Shapiro, G. Shvets, J. R. Sirigiri and R. J. Temkin, \textit{Opt. Lett.} \textbf{31}, 2051 (2006).
\bibitem{Garc:2005} F. J. Garc\'{\i}a de Abajo and J. J. S\'{a}enz, Phys. Rev. Lett. \textbf{95}, 233901 (2005).
\bibitem{Marques:2002} R. Marques, F. Medina, and R. Rafii-El-Idrissi, \textit{Phys. Rev. B} \textbf{65}, 144440 (2002).
\bibitem{Xudong:2005} Xudong Chen, Bae-Ian Wu, Jin Au Kong, and Tomasz M. Grzegorczyk , \textit{Phys. Rev. E} \textbf{71}, 046610 (2005).
\bibitem{Schurig:2006} D. Schurig, J. J. Mock, and D. R. Smith, \textit{Appl. Phys. Lett.} \textbf{88} 041109 (2006).
\bibitem{Dmitriev:2000} V. Dmitriev, \textit{Prog. Electromagn. Res.} \textbf{28}, 43 (2000). 
\bibitem{Ponti:2001} S. Ponti, C. Oldano, and M. Becchi, \textit{Phys. Rev. E} \textbf{64}, 021704 (2001).
\bibitem{Ponti:2002} S. Ponti, J. A. Reyes and C. Oldano, \textit{J. Phys.: Cond. Matter} \textbf{14}, 10173 (2002).
\bibitem{Juan:2004} Juan D. Baena, Ricardo Marqu\'{e}s, Francisco Medina, and Jes\'{u}s Martel, \textit{Phys. Rev. B} \textbf{69} 014402 (2004).
\bibitem{Belov:2005} P. A. Belov, and C. R. Simovski, \textit{Phys. Rev. E} \textbf{72}, 026615 (2005).
\bibitem{Tretyakov:2003} S. A. Tretyakov, \textit{IEEE Trans. Antennas Propag.} \textbf{51} 2652 (2003).
\bibitem{Mahan:2006} G. D. Mahan, \textit{Phys. Rev. B} \textbf{74}, 033407 (2006).
\bibitem{Boardman:2006} A. D. Boardman and K. Marinov, \textit{Phys. Rev. B} \textbf{73}, 165110 (2006).
\bibitem{Tretyakov_book:2003} S. Tretyakov, Analytical Modeling in Applied Electromagnetics (Artech House, Boston 2003).
\bibitem{Koschny:2005} Th. Koschny, et.al., \textit{Phys. Rev. B} \textbf{71}, 245105 (2005).
\bibitem{Kempa:2005} K. Kempa, R. Ruppin and J. B. Pendry, \textit{Phys. Rev. B} \textbf{72}, 205103 (2005). 
\end{thebibliography}
\end{document}